\documentclass[aps,prd,reprint,amsfonts,amssymb,amsmath,showpacs,preprintnumbers, letterpaper,nofootinbib,urlcolor=black,linkcolor=black]{revtex4-1}

\usepackage{array}

\usepackage{cancel}
\usepackage{braket,bm}
\usepackage[pass]{geometry} 
\usepackage{color}
\usepackage{mathrsfs,mathtools}
\usepackage{graphicx}
\usepackage{bbm}
\usepackage{enumerate}
  \DeclareMathAlphabet{\mathpzc}{OT1}{pzc}{m}{it}
\usepackage{subfigure}
\usepackage{graphicx}  
\usepackage{soul}
\usepackage{comment}

\usepackage[colorlinks,citecolor=black]{hyperref}

\newcommand{\G}{{G}}

\def\d{\mathrm{d}}
\def\D{\mathcal{D}}

\def\e{\mathrm{e}}
\def\i{\mathrm{i}}

\begin{document}

\title{Schwinger effect 
and false vacuum decay
as quantum-mechanical tunneling of a relativistic particle}
\author{Wen-Yuan Ai}
\email{wenyuan.ai@uclouvain.be}

\author{Marco Drewes}
\email{marco.drewes@uclouvain.be}

\affiliation{Centre for Cosmology, Particle Physics and Phenomenology,\\ Université catholique de Louvain, Louvain-la-Neuve B-1348, Belgium}

\preprint{CP3-20-24}

\begin{abstract}
We present a simple and intuitive description of both, the
Schwinger effect 
and false vacuum decay through bubble nucleation, 
as tunneling problems in one-dimensional relativistic quantum mechanics. Both problems can be described by an effective potential that depends on a single variable of dimension length, which measures the separation of the particles in the Schwinger pair, or the radius of a bubble for the vacuum decay. We show that both problems can be described as tunneling in one-dimensional quantum mechanics if one interprets this variable as the position of a relativistic particle with a suitably defined effective mass. The same bounce solution can be used to obtain reliable order of magnitude estimates for the rates of Schwinger pair production and false vacuum decay.
\end{abstract}


\maketitle

\section{Introduction}
\label{sec:intro}
The Schwinger effect~\cite{Schwinger:1951nm} and phase transitions through bubble nucleation~\cite{Kobzarev:1974cp,Frampton:1976kf,Coleman:1977py,Callan:1977pt} are both nonperturbative effects in which a metastable state decays into an energetically favorable configuration.
In the case of the Schwinger effect (see Ref.~\cite{Gelis:2015kya} for a review) the metastable state is a very strong static electric field that decays through spontaneous production of electron-positron pairs~\cite{Schwinger:1951nm,Sauter:1931zz,Heisenberg:1935qt}.
Phase transitions can be modeled in scalar quantum field theory 
as the decay of a metastable state known as ``false vacuum," where the scalar field represents an order parameter for the transition. 

In both problems, there is potential energy $U(x)$ that can be characterized by a single variable $x$ of dimension length, which measures the separation of the particles in the Schwinger pair or the radius of a bubble for the vacuum decay. The strong electric field in empty space and the false vacuum can both be identified with local minima in $U(x)$ at $x=0$, and in both problems there is a critical distance $x_c > 0$ with $U(x_c) = U(0)$ beyond which the potential energy is smaller than $U(0)$. For the Schwinger pair, this occurs because the rest masses of the electron-positron pair are overcompensated by their potential energy in the external field for $x>x_c$. False vacuum decay occurs when the bubble radius is large enough that the volume-dependent energy difference between the two phases exceeds the surface-dependent energy needed to form the bubble wall. The decay of the metastable states can then be viewed as a quantum mechanical tunneling from $x=0$ to $x=x_c$, and as usual for tunneling, the decay rate per unit volume is exponentially suppressed.
It is common to express this rate as \begin{align}
\label{eq:tunnelingrate2}
\varGamma/V=A\, \e^{-B},
\end{align}
where the coefficient $B$ is approximately a polynomial in the coupling constants in the theory, reflecting the nonperturbative nature of the decay.
In the present work, we adopt a very simple picture and show that we can correctly reproduce the known expression for $B$ and obtain an order of magnitude $A$ in both problems by mapping them onto the tunneling of a relativistic particle in one-dimensional quantum mechanics. Here  $x$ is interpreted as the position of the particle, and the potential energy $U(x)$ has to be complemented by a kinetic energy with a suitably defined effective mass. In this picture, the same bounce solution can be used to compute $B$ in both problems. This makes the analogy between these two phenomena very explicit.

Before deriving our results, we briefly recollect some of the main results from previous computations that we compare to in Sec.~\ref{sec:history}. In Sec.~\ref{sec:schwinger}, we show that Schwinger effect can be studied as quantum-mechanical tunneling for a relativistic particle. In Sec.~\ref{sec:FVD}, we show that this picture on Schwinger effect is analogous to false vacuum decay in the thin-wall regime. Section~\ref{sec:conclusions} is left for conclusions and discussions. To be self-contained,  a brief review of the Callan-Coleman method on quantum tunneling 
is given in the Appendix.

\section{A simple model for Schwinger effect}

\subsection{Schwinger effect in other approaches}\label{sec:history}

The original derivation of Schwinger effect was given in quantum electrodynamics (QED) and based on a calculation of the vacuum to vacuum transition (or vacuum persistence) amplitude in the presence of an external static electric field. This amounts to computing the QED effective action whose imaginary part can be related to the decay of the vacuum.\footnote{Here the vacuum denotes the QED vacuum in the presence of an external classical electric field. If backreaction is taken into account this corresponds to the decay of the electric field, which could, e.g.,~be described by treating the field as a quantum object. We shall distinguish this phenomenon from false vacuum decay where we have multiple vacua in the absence of any external field.} 
The widely used tool for this computation is the Schwinger proper time method~\cite{Schwinger:1951nm}. Inspired by string theory, a similar method to Schwinger proper time, called the worldline formalism, has been applied to the study of Schwinger effect in inhomogeneous external fields~\cite{Dunne:2005sx,Dunne:2006st}.
Schwinger effect can also be studied in the canonical way. In the time-dependent gauge for the electromagnetic potential, one studies the Bogolyubov transformation~\cite{Bogolyubov:1958km} between the in-modes and out-modes
in the presence of a time-dependent external electromagnetic potential~\cite{Brezin:1970xf,Narozhnyi:1970uv,Gitman:1977ne,Soffel:1982pm,Ambjorn:1982bp,Gavrilov:1996pz,Gavrilov:2006jb,Tanji:2008ku}. Setting the in-state to be the vacuum, the negative-frequency coefficient in the Bogolyubov transformation gives the number of the produced particles in the out-state. This method can also be used to study Schwinger effect in non-Abelian gauge theories~\cite{Domcke:2018gfr,Domcke:2019qmm}.

The rate of pair production in all approaches is consistently determined to be proportional to $\exp(-\pi m^2/qE)$.
The nonanalytic dependence on the coupling $q$ indicates the nonperturbative nature of this effect. 
For comparison, the pair production rate was directly computed by Nikishov~\cite{Nikishov:1970br} and is given as
\begin{align}\label{Nikishov}
\varGamma=\frac{(qE)^2}{4\pi^3}\e^{-\pi m^2/qE},
\end{align}
where $m$ and $q$ are the positron mass and charge, respectively. $E$ is the magnitude of the static electric field.  One may also consider the simpler $1+1$-dimensional problem where the pair production rate is 
\begin{align}
\label{eq:pairproductionrate2}
\varGamma=\frac{qE}{2\pi}\e^{-\pi m^2/qE}.
\end{align}

The tunneling interpretation on the basis of a potential energy and the analogy between both problems are of course well known.
As was first pointed out by Brezin and Itzykson~\cite{Brezin:1970xf}, the exponential dependence in the pair production rate is reminiscent of quantum tunneling suppression. This is usually illustrated with the qualitative picture of Dirac sea, where a negative-frequency state tunnels to a positive-frequency state, leading to pair production. 
The tunneling analogy has also been used to estimate thermal corrections to Schwinger effect~\cite{Brown:2015kgj} and in the context of the Dirac-Born-Infeld brane tunneling~\cite{Brown:2007zzh,Brown:2007vha}. In Refs.~\cite{Garriga:2012qp,Garriga:2013pga,Frob:2014zka}, the authors use Schwinger effect to study conceptual subtleties appearing in bubble nucleation.
A quantitative implementation of the tunneling picture perhaps comes from the canonical method in the space-dependent gauge. In this case, because the electromagnetic potential is time independent, one instead ends up with a static Schr\"{o}dinger-like equation with a potential barrier from Klein-Gordon equation (in scalar QED) or Dirac equation. With certain assumptions for the interpretations of the incoming and outgoing waves, one can recover the pair production rate or the Schwinger formula for the vacuum decay rate~\cite{Nikishov:1970br,Casher:1978wy,Olsen:1981xn,Srinivasan:1998ty,Kim:2000un,Kim:2003qp}. 
Our analysis differs from this approach because it relies on neither the Klein-Gordon nor the Dirac equation.

\subsection{Schwinger effect as quantum-mechanical tunneling}
\label{sec:schwinger}

From a quantum physics viewpoint, very strong electric fields represent an unstable state that must decay to the true ground state.
Schwinger pair production is the dominant process that drives this decay.\footnote{In many works in the literature, the vacuum decay rate and pair production rate are identified with each other, while in fact the latter is simply the leading contribution to the former. Processes with more particles in the final state give subdominant contributions~\cite{Cohen:2008wz}.} We now derive the pair production rate in the effective quantum-mechanical-tunneling picture as described in the Introduction. We start with the simpler $1+1$-dimensional Schwinger effect. Let the positions of the positron and the electron be $x_1$ and $x_2$ respectively. Then the classical energy for a particle pair is 
\begin{align}\label{PairEnergy}
\mathcal{E}=\frac{m}{\sqrt{\
1-\dot{x}_1^2}}+\frac{m}{\sqrt{\
1-\dot{x}_2^2}}-qE(x_1-x_2),
\end{align}
with $\dot{x}_i=\partial_t x_i$, and we have neglected the Coulomb force between the two particles. Choosing the frame of mass center with $x_1=X+x$ and $x_2=X-x$, we obtain
\begin{align}
\label{eq:energypair}
\mathcal{E}=\frac{2m}{\sqrt{\
1-\dot{x}^2}}-2qEx.
\end{align}
To study the tunneling problem, let us first look at the energy for a particle pair at rest which is $\widetilde{U}_{\rm eff}(x)=2m-2qEx$. To take into account the fact that there are no particles initially, we modify the potential as $U_{\rm eff}(x)=\widetilde{U}_{\rm eff}(x)-2m\tilde{\delta}(x)$ where $\tilde{\delta}(x)$ equals one for $x=0$ and zero otherwise so that $U_{\rm eff}(x=0)=0$. For $x\neq 0$, there is another point $x_c\equiv m/(q E)$ at which the potential is zero, {i.e.,} 
$U_{\rm eff}(x_c)=0$. For $0 < x < x_c$, energy conservation cannot be satisfied. 
Schwinger effect can be described by the quantum tunneling of a relativistic particle from $x=x_0=0$ to $x=x_c$. Equation~\eqref{eq:energypair} can be written as 
\begin{equation}\label{MinkowskiEoM}
\dot{x}^2 = 1 - \left(
x/x_c  + \mathcal{E}/(2m) 
\right)^{-2}.    
\end{equation}

The Hamiltonian corresponding to Eq.~\eqref{eq:energypair} can be obtained from the following action:
\begin{align}
\label{action}
S=\int\d t\, \left[-2m\sqrt{1-\dot{x}^2}+2qEx+2m\tilde{\delta}(x)\right],
\end{align}
where $2m\tilde{\delta}(x)$ has been included such that the energy vanishes for $x=0$. This term does not affect the dynamics whenever $x\neq 0$. 
In principle, tunneling through arbitrarily shaped barriers in quantum mechanics can be described in terms of wave mechanics. However, the kinetic term in the action \eqref{action} is not canonical, so that known solutions of the Schr\"odinger cannot be used. We therefore resort to the principle of minimal action to compute the pair production rate $\varGamma$, taking advantage of the fact that the path integral for quantum mechanical amplitudes is dominated by trajectories near the classical solution. In the case of tunneling, this is somewhat problematic because there are no classical trajectories, which can directly be verified by noticing that \eqref{MinkowskiEoM} has no real solutions starting at $x=0$ for $\mathcal{E}<2m$; for the case $\mathcal{E}=0$ under consideration here, valid (real) classical solutions only exist for $x\geq x_c$. 
This problem can be circumvented by considering the analytic continuation to Euclidean space $t\rightarrow -\i\tau$ and following the standard Callan-Coleman method~\cite{Callan:1977pt}. We give a brief summary of this method in the Appendix. 
In this approach, the strong field can be physically interpreted as an unstable bound state. Its decay rate $\varGamma$ can be obtained from the imaginary part of the
(approximate) eigenvalue $E_0$ of the Hamiltonian corresponding to the lowest bound state in the $\tilde{\delta}(x)$-potential, which would be stable for $E=0$ or $q=0$.
This is in analogy to the ``width" of unstable particles in particle physics, where $q$ takes the role of a coupling constant that is responsible for the decay.
A subtlety arises due to the fact that we have ignored the center of mass coordinate $X$ in the action \eqref{action}.
We account for this by first defining the transition rate $\widetilde{\varGamma}$ for $X=0$, which we later relate to the full rate $\varGamma$ that takes into account the fact that the transition can happen anywhere in space. The decay rate $\widetilde{\varGamma}$ is given as 
\begin{align}
\label{Gamma1}
\widetilde{\varGamma}=-2\;{\rm Im} E_0=\lim_{\mathcal{T}\rightarrow\infty}\frac{2}{\mathcal{T}}{\rm Im}(\ln Z^E[\mathcal{T}])
\end{align}
with
\begin{align}
\langle x_0|\e^{-H\mathcal{T}}|x_0\rangle=\int_{x(\tau=\pm\mathcal{T})=x_0}\D x\, \e^{-S_E}\equiv Z^E[\mathcal{T}],
\end{align}
where $\mathcal{T}$ is the amount of Euclidean time and $S_E$ is the Euclidean action,
\begin{align}
\label{eq:Eucaction-schwinger}
S_E=\int\d\tau\,\left[2m\sqrt{1+\left(\frac{\d x}{\d\tau}\right)^2}-2qEx-2m\tilde{\delta}(x)\right].
\end{align}
$|x_0\rangle$ is the state with the particle at $x=x_0=0$. The path integral is to be taken over all trajectories that begin and end at $x=x_0$ at $\tau=\pm\mathcal{T}$. Formally, this corresponds to  the {\it Euclidean} transition amplitude from $x=0$ to itself, but the expression is only needed as an auxiliary tool here.

\begin{figure}[h]
\centering
\includegraphics[scale=0.5]{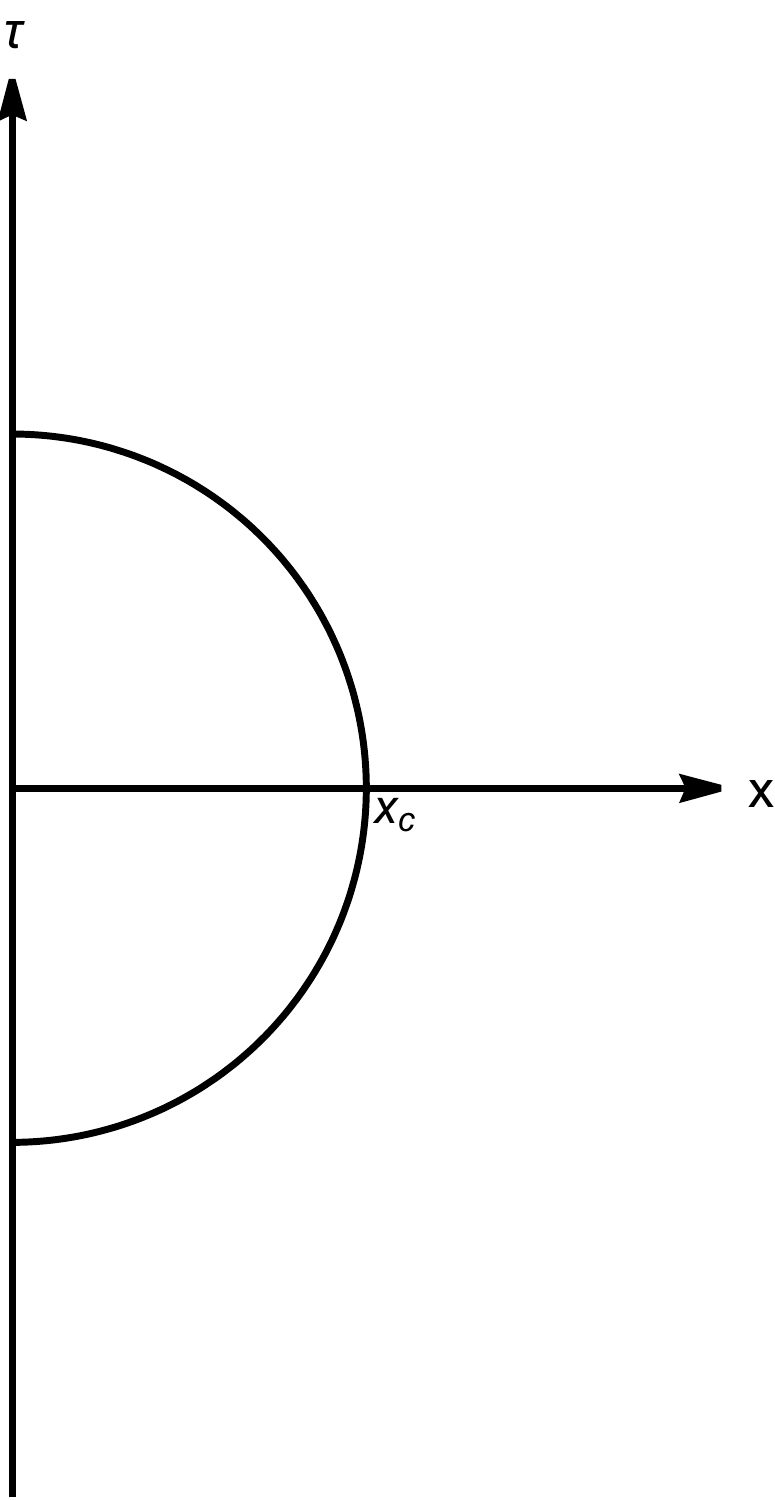}
\ \ \ \ \ \ \ \ \ \ \ \ \ 
\includegraphics[scale=0.5]{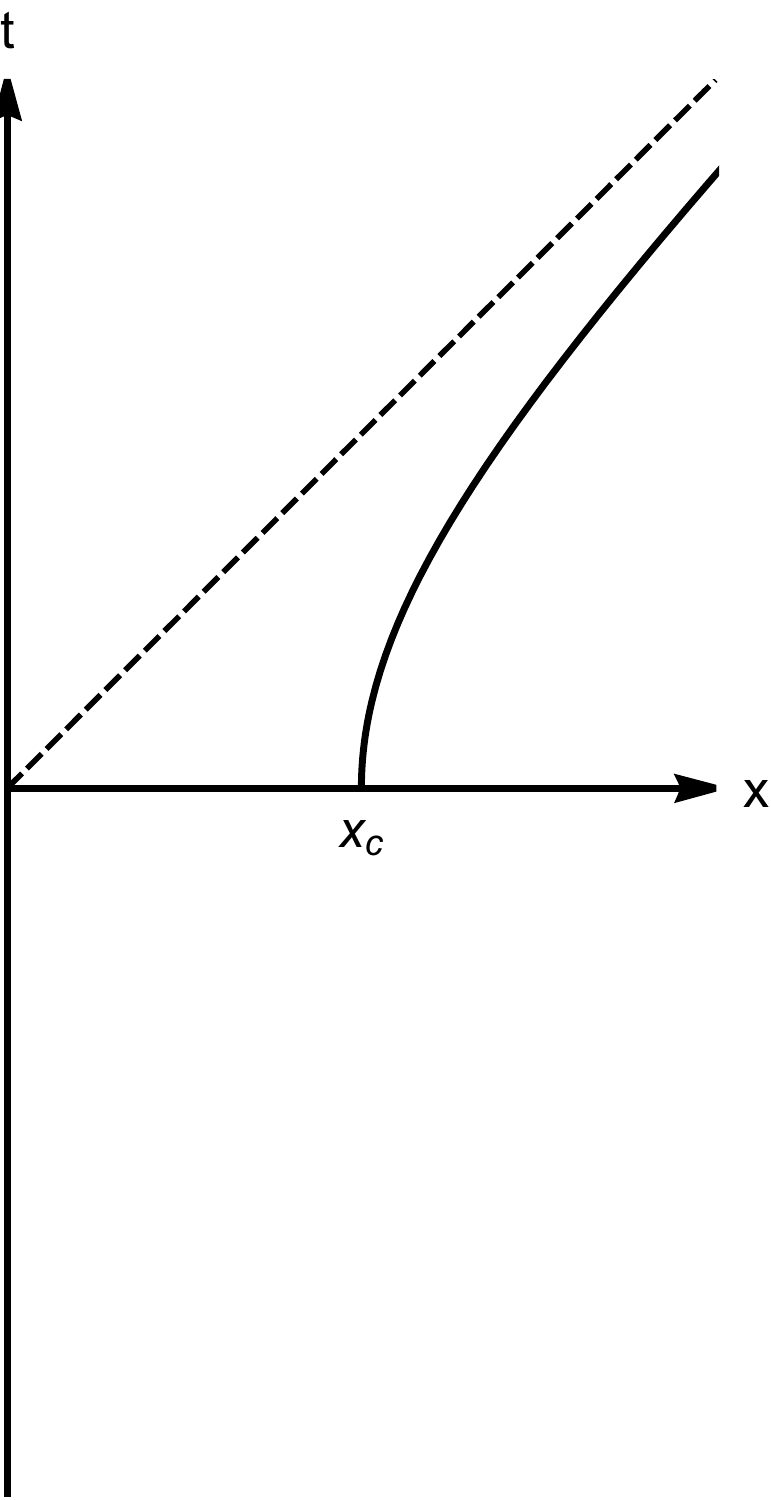}
\caption{On the upper panel displays the instanton solution $x_B(\tau)$ in \eqref{eq:motion}. 
The lower panel shows its analytic continuation to Minkowski space.
If can be interpreted as the creation of a Schwinger pair that is created with separation $x_c$ at $t=0$. For $t>0$, the particles are accelerated by the electric field to velocities that approach the speed of light, and the separation $x$ grows rapidly. 
}
\label{fig:bounce}
\end{figure}

One can estimate the path integral using the method of steepest descent. In contrast to the Minkowski space equation \eqref{MinkowskiEoM}, its analytic continuation in Euclidean space for $\mathcal{E}=0$,
\begin{align}
\label{eq:eom-schwinger}
\left(\frac{\d x}{\d\tau}\right)^2=-1+\frac{x_c^2}{x^2},
\end{align}
has, besides the trivial solution $x_F(\tau)\equiv 0$,\footnote{
The trivial solution $x_F(\tau)$ cannot be derived by Eq.~\eqref{eq:eom-schwinger} which is valid only for $x\neq 0$ because we have omitted the $\tilde{\delta}(x)$-term. Including the $\tilde{\delta}(x)$-potential, we see that $x_F(\tau)$ conserves energy.} 
an instanton solution $x_B(\tau)$ (named as bounce) for the boundary condition $x(\tau\to\pm\mathcal{T})=0$,\footnote{We have assumed that, without loss of generality, the center of the bounce is at $\tau=0$.
Further, we should note that only the real part of $x_B(\tau)$ is physical; we have only done an analytic continuation in the time variable.
} 
\begin{align}
\label{eq:motion}
x_B(\tau)&=\sqrt{x_c^2-\tau^2}\ \ \ {\rm for}\ -x_c\leq\tau\leq x_c, \notag\\ 
x_B(\tau)&=0\ \ \ {\rm for\ others}.
\end{align}
The solution \eqref{eq:motion} is shown in Fig.~\ref{fig:bounce}.
Its continuation to Minkowski space has a simple physical interpretation: the particle-antiparticle pair is spontaneously created at $t=0$ with vanishing velocity, but then accelerated due to the electric field, and their velocity asymptotically approaches the speed of light, cf.~Fig.~\ref{fig:bounce}.
In the Appendix, we briefly summarize how to estimate the path integral in the expression \eqref{Gamma1} by expanding the Euclidean path integral around the trivial solution $x_F(\tau)$ and the bounce solution $x_B(\tau)$. After including the possibility of multiple subsequent bounces in the so-called dilute-gas approximation, one finally finds that the coefficient $B$ is simply given by the difference in the actions associated with the classical trajectories $x_F(\tau)$ and $x_B(\tau)$,
\begin{equation}\label{BfromS}
B=S_E[x_B]-S_E[x_F]=S_E[x_B],
\end{equation}
where we have used $U_{\rm eff}(x=0)=0$.
Substituting Eq.~\eqref{eq:motion} into Eq.~\eqref{BfromS} with Eq.~\eqref{eq:Eucaction-schwinger}, we obtain
\begin{align}\label{Meinholz}
B=\frac{\pi m^2}{qE},
\end{align} 
in agreement with the results \eqref{Nikishov} and \eqref{eq:pairproductionrate2} derived from the field theory.

While the exponent $B$ in \eqref{eq:tunnelingrate2} can be computed from the action along the classical trajectories \eqref{BfromS} alone, the prefactor $A$ depends on quantum fluctuations. 
For small (Gaussian) fluctuations, these can be obtained from a functional Taylor expansion  around the classical path, cf.~\eqref{ZE_exp_formula},
which yields the fluctuation operator
\begin{align}
\label{eq:fluct-operator}
    \G(\tau_1,\tau_2)=\frac{\delta^2 S_E}{\delta x(\tau_1)\delta x(\tau_2)}.
\end{align} 
This amounts to computing the functional determinants of \eqref{eq:fluct-operator} evaluated for the bounce and false vacuum, cf.~Eq.~\eqref{FunctionalDeterminants}. This is rather involved in practice, and instead of a rigorous computation, we obtain an estimate based by exploiting the observation made in the original paper \cite{Callan:1977pt} that there is a zero mode in the eigensystem of $\G$ for each spacetime symmetry. 
 The action \eqref{action} is time translation invariant; the pair production can happen anytime. The zero mode corresponding to the time-translation symmetry gives a contribution $\mathcal{T}\sqrt{B/2\pi}$ to $A$. 
 The fact that the pair creation can happen anywhere in space is not captured by the action \eqref{action} and has to be fixed by hand by introducing the analogous factor $V\sqrt{B/2\pi}$,
 \begin{align}
     \varGamma=\left(V\sqrt{\frac{B}{2\pi}}\right)\widetilde{\varGamma}.
 \end{align}
In total, we have a factor $V\mathcal{T}(B/2\pi)$ in $\varGamma$. The factor $V\mathcal{T}$ will be canceled out by the $\mathcal{T}$ and $V$ appearing in Eqs.~\eqref{eq:tunnelingrate2} and~\eqref{Gamma1}. 
The contributions from all other modes (including the negative mode) are difficult to calculate; see, e.g., Ref.~\cite{Ai:2019fri}. By dimensional analysis, we know $A$ has dimension two in mass. Since the characteristic scale in the tunneling process is $x_c$, we thus estimate $A$ as
\begin{align}
A\approx \frac{B}{2\pi}\frac{1}{x_c^2}=\frac{qE}{2}.
\end{align}
Comparing with Eq.~\eqref{eq:pairproductionrate2}, we find that this gives a correct order of magnitude estimate.
The somewhat surprising success of our very simple approach can be better understood by noting that the action \eqref{action} is similar to an action that can be derived by using the so-called Schwinger proper time representation~\cite{Corradini:2015tik}. However, we emphasize that we do not need the field theoretical framework on which applications of this method to the Schwinger effect are usually based~\cite{Dunne:2005sx,Gelis:2015kya}, but we could simply guess \eqref{action} from \eqref{PairEnergy}
based on physical intuition.

The analysis in $3+1$-dimensional spacetime is similar since only the spatial direction with nonvanishing external electric field is most relevant. The main difference is that now we have four translation symmetries which contribute $V\mathcal{T}(B/2\pi)^2$ in $Z_B^E[\mathcal{T}]$, cf.~\eqref{Aestimate}, where  $V$ is now the volume for three-dimensional space. 
Then, we estimate $A$ as
\begin{align}
A\approx\left(\frac{B}{2\pi}\right)^2\frac{1}{x_c^4}=\frac{(qE)^2}{4\pi^2}.
\end{align}
A more careful comparison for $A$ in our method and other field-theoretical methods is left for future work.

\subsection{Schwinger effect in spatially inhomogeneous electric fields}

Before ending this section, we note that our method can also be generalized to cases with inhomogeneous external fields. For example, in the $1+1$-dimensional case, this means that the simple function $-2qEx$ may be replaced by a more complicated function $f(x)$ and the spatial-translation symmetry may be broken by the external field already. In that case, we shall study the total production rate for a specific event instead of the rate per unit volume. 

As an example for illustration, we consider a Sauter-type electric field~\cite{Sauter:1931zz} in $1+1$-dimensional case, 
\begin{equation}\label{Sauter}
E(x)=E{\rm sech}^2(kx). 
\end{equation}
We assume that the particle production is most efficient at the origin $x=0$ because that is where the field is the strongest.
This electric field is symmetric with respect to the origin  and as a result, one expects that the particle pair is likely nucleated with positions being at $x$ and $-x$. 
The energy of the system for a given separation $x$
is obtained by simply integrating $E(x)$ over $x$,
\begin{align}\label{SauterEnergy}
\mathcal{E}=\frac{2m}{\sqrt{1-\dot{x}^2}}-\frac{2qE}{k}\tanh(kx).
\end{align}
From the above equation, one immediately knows that the critical value of $x$ for the nucleated pair is
\begin{align}
\tanh(k x_c)=\frac{mk}{qE} = k x_c \equiv \gamma.
\end{align}
In order to have real finite $x_c$, we must require $\gamma<1$, which amounts to $E > m k/q$, i.e., there is a minimal field strength $E$ that is required for the effect to occur. 
Recalling that $1/k$ in \eqref{Sauter} characterizes the spacial extension of the region where the electric field is present, we can also read this condition as $1/k > x_c$, which simply means that there is no pair production if the region where the field is present is smaller than the critical distance $x_c$.

The corresponding Euclidean action now is
\begin{align}
S_{\rm E}=\int\d\tau\left[2m\sqrt{1+\left(\frac{\d x}{\d\tau}\right)^2}-\frac{2qE}{k}\tanh(kx)-2m\tilde{\delta}(x)\right],
\end{align}
and the equation of motion is 
\begin{align}\label{EoMinhomogeneous}
\left(\frac{\d x}{\d\tau}\right)^2=-1+\frac{\gamma^2}{\tanh^2(kx)},
\end{align}
which has the same form as \eqref{eq:eom-schwinger} with the replacement $x\to \tanh(kx)$.
The above equation has an analytic bounce solution
\begin{align} 
x_B(\tau)=\frac{1}{k}{\rm arcsinh}\left(\sqrt{\frac{\gamma^2-\sin^2\left(\sqrt{1-\gamma^2}k|\tau|\right)}{1-\gamma^2}}\right),
\end{align}
for 
\begin{align}
|\tau|\leq \frac{{\rm arcsin(\gamma)}}{\sqrt{1-\gamma^2}k},
\end{align}
and $x_B(\tau)=0$ for others. Substituting the bounce solution into the Euclidean action, one obtains the semiclassical tunneling rate
\begin{align}\label{Binhomogeneous}
B
=\frac{m^2\pi}{qE}
\frac{2}{1+\sqrt{1-\gamma^2}}
.
\end{align}
The result \eqref{Binhomogeneous} is physically intuitive if we interpret $1/\gamma \propto 1/k$ as a measure for the extension of the region where the electric field is present.
For $\gamma\ll1$, this region is much larger than the critical distance $x_c$, and we recover the result \eqref{Meinholz} for a homogeneous field. This makes sense because the function \eqref{Sauter} is approximately constant over distance $x_c$ near the origin.
For $\gamma>1$, the region where the field is present is smaller than the critical distance $x_c$, and no tunneling happens. Note that in this regime the rate is strictly zero and not just exponentially suppressed. 
 In other words, for any value of $k$, there exists a critical field strength $E_{\rm crit}=mk/q$ that is needed for pair creation to happen. For $E<E_{\rm crit}$, the pair creation rate vanishes, at least within the approximations made here. This is very different from the homogeneous field case where the pair creation rate is nonzero even for arbitrarily small $E$, but simply becomes exponentially suppressed.
This difference can be understood physically. In the homogeneous field case, the absolute value of the potential energy in \eqref{eq:energypair} is unbound, it grows linearly with $x$, and always exceeds the pairs' rest energy $2m$ for sufficiently large separation. In practice, pair creation does not happen for $E\ll \pi m^2/q$ because the distance $x_c$ that the system would have to tunnel becomes too large, but in principle it is possible.
In contrast, for the localized field \eqref{Sauter}, the absolute value of the potential energy in Eq.~\eqref{SauterEnergy} is bounded from above and cannot exceed $2qE/k$. For $E<E_{\rm crit}$, this maximal value is smaller than $2m$.

Our result \eqref{Binhomogeneous} is
in agreement with the result given in Ref.~\cite{Dunne:2005sx}. This shows that our simple method can be used to treat inhomogeneous electric fields.
It would further be interesting to see whether our simple method can be further generalized to reproduce known results for the Schwinger effect in curved spacetime (see, {e.g.},~\cite{Garriga:1994bm,Kim:2008xv,Chen:2012zn,Frob:2014zka,Cai:2014qba,Kobayashi:2014zza,Fischler:2014ama,Stahl:2015gaa,Bavarsad:2016cxh,Hayashinaka:2016qqn,Gorbar:2019fpj,Sobol:2020frh,Chen:2020mqs}) or non-Abelian fields (see, {e.g.,}~\cite{Domcke:2018gfr}). 
This, however, clearly goes beyond the scope of the present paper.

\section{Analogy with False Vacuum Decay}
\label{sec:FVD}

The above quantum-mechanical-tunneling picture on Schwinger effect can be used to draw a close analogy to false vacuum decay in quantum field theory, which can be mapped onto a simple tunneling problem in one-dimensional quantum mechanics in the same way. To see it, we recall the most important aspects of false vacuum decay in the thin-wall regime. 

\subsection{Brief review of false vacuum decay in field theory}

\begin{figure}[h]
\centering
\includegraphics[scale=0.55]{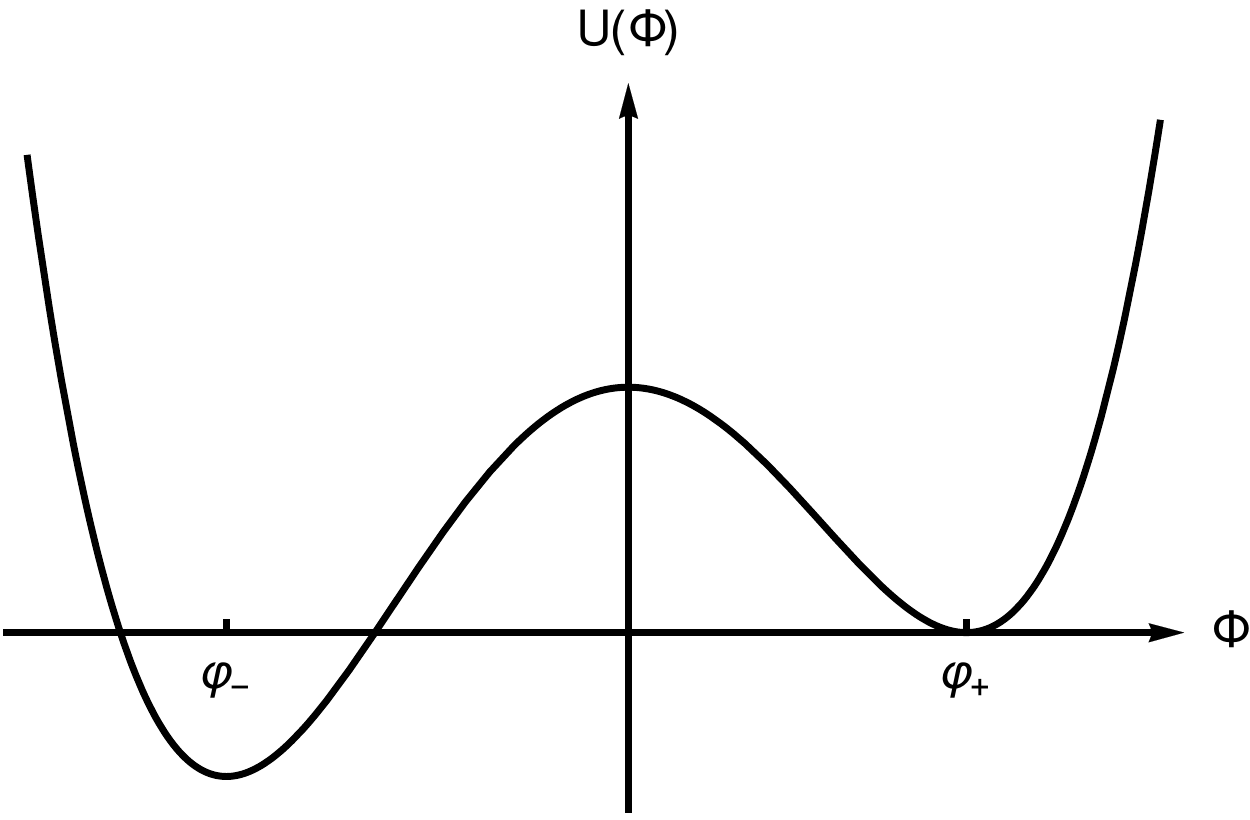}
\caption{The classical  potential $U(\Phi)$ for a theory with a false vacuum.}
\label{fig:potential}
\end{figure}
We consider a scalar field theory with the Euclidean action
\begin{align}
\label{Callan-Coleman}
S_E=\int\d^4x \left[\frac{1}{2}(\partial_\mu\Phi)^2+U(\Phi)\right].
\end{align}
The quantum-mechanical picture that we develop in Sec.~\ref{vac:QM} holds for a wide class of potentials. 
For the sake of definiteness we choose the same potential as in the work by Coleman to make connection to the known results in quantum field theory,
$  U(\Phi)=-\frac{1}{2}\mu^2\Phi^2+\frac{1}{3!}g\Phi^3+\frac{1}{4!}\lambda\Phi^4+U_0$,
with $\mu^2,g,\lambda$ are positive real parameters.
$U(\Phi)$ exhibits a metastable minimum, as sketched in Fig.~\ref{fig:potential}.
The false and true vacua can be understood as local minima in the energy functional associated with different configurations of the background field $\varphi\equiv\langle\Phi\rangle$. For simplicity, we choose the constant $U_0$ such that $U(\varphi_+)=0$ at the false vacuum configuration $\varphi_+$. The equation of motion for $\varphi$ then reads
 \begin{align}\label{CrocodileBatidaSpecial}
-(\partial_t^2 - \nabla^2)\varphi+U'(\varphi)=0.
\end{align}
The two configurations $\varphi_+$ and $\varphi_-$ must be spatially homogeneous and isotropic to avoid gradient energies. There are no classically allowed trajectories in field space that connect them, but tunneling through the barrier in Fig.~\ref{fig:potential} is possible at the quantum level. Using the Coleman-Callan method, the transition can again be described by performing a continuation to Euclidean space.

The decay rate per unit volume has the form as in Eq.~\eqref{eq:tunnelingrate2}, 
and the semiclassical suppression factor is again given by
\begin{align}
B=S_E[\varphi_B],
\end{align}
where $\varphi_B$ is the bounce. The bounce satisfies the equation of motion \eqref{CrocodileBatidaSpecial} in Euclidean time $\tau$ and the radial coordinate $r$
with the boundary conditions $\varphi|_{\tau\rightarrow \pm\infty}=\varphi_+$ and $\dot{\varphi}|_{\tau=0}=0$, where $'$ and the dot denote the derivatives with respect to the field $\varphi$ and $\tau$, respectively. For the theory we consider, it can be shown that the bounce has $O(4)$ symmetry~\cite{Coleman:1977py}.
The physical reason is that 
the false vacuum decay happens via bubble nucleation, and spherical bubbles are for energetic reasons the most likely configuration. 
Therefore, the equation of motion can be written as
\begin{align}
\label{eom}
-\frac{\d^2\varphi}{\d\rho^2}-\frac{3}{\rho}\frac{\d\varphi}{\d \rho}+U'(\varphi)=0,
\end{align}
where $\rho^2=r^2+\tau^2$, with the boundary conditions $\varphi|_{\rho\rightarrow\infty}=\varphi_+$ and $\d\varphi/\d\rho|_{\rho=0}=0$. 
As did for Schwinger effect, we will only concern about $B$. For the calculations of $A$ in false vacuum decay, see, {e.g.,} Refs.~\cite{Callan:1977pt,Isidori:2001bm,Branchina:2014rva,Garbrecht:2015oea,Andreassen:2017rzq,Chigusa:2018uuj,Garbrecht:2018rqx,Ai:2018guc,Ai:2020sru} and especially Ref.~\cite{Ai:2019fri} for a comparison of different methods. In the thin-wall approximation, the damping term in Eq.~\eqref{eom} can be neglected, and the solution is given by a ``kink."
For case $g^2/\mu^2\ll \lambda$ and $\varphi_-\simeq -\varphi_+$,
one can find an analytic solution of the form
\begin{equation}\label{kink}
\varphi = \varphi_+\tanh(\gamma (\rho-R_c)).
\end{equation}
The kink connects  the false vacuum outside of the bubble with the true vacuum inside.
 $\gamma$ and $R_c$ are parameters that depend on the details of the model and can be expressed in terms of the parameters in the potential, see {e.g.,} Refs.~\cite{Garbrecht:2015oea,Ai:2018guc}. 
In physical terms, $R_c$ is the radius of a critical bubble and $1/\gamma$ is a measure for the thickness of the bubble wall. 

We denote the energy difference between the false vacuum and true vacuum as $\epsilon$, {i.e.}, $\epsilon=U(\varphi_+)$.
 Outside the wall, the Euclidean action $B_{\rm outside}=0$. Inside the wall,
\begin{align}
B_{\rm inside}=-\frac{\pi^2}{2}R_c^4\epsilon.
\end{align}
The region near the wall contributes
\begin{align}
\label{surfacetension}
B_{\rm wall}&=2\pi^2R_c^3\int_{R_c-\delta}^{R_c+\delta}\d\rho\left[\frac{1}{2}\left(\frac{\d\varphi}{\d\rho}\right)^2+U(\varphi)\right]\notag\\
&\equiv 2\pi^2R_c^3\sigma,
\end{align}
where $\delta$ is a large enough number compared with the characteristic scale of the bubble-wall 
width and we have defined the surface tension $\sigma$ of the bubble wall. 
 
The radius of the critical bubble $R_c$ is given by the stationary point of $B$, $\d B/\d R_c=0$. We thus have $R_c=3\sigma/\epsilon$ and the decay suppression $B=27\pi^2\sigma^4/2\epsilon^3$. Due to the $O(4)$ symmetry, the bounce has the profile shown in Fig.~\ref{fig:bounce} with the replacement $(x,x_c)\to(\rho,R_c)$,
where the solid circle represents the bubble wall
separating the false vacuum and the true vacuum defined by $\rho=R_c$. 
The analytic continuation $\tau\rightarrow \i t$ of this condition into Minkowski space reads $r^2 - t^2 = R_c^2 $. 
Solving for $r$ yields an expression for the radius $R(t)$ of an expanding bubble that nucleates at time $t=0$ with radius $R_c$ and then expands, $R(t) = (R_c^2 +  t^2)^{1/2}$. It can be seen in Fig.~\ref{fig:bounce} with the replacement $(x,x_c)\to(R,R_c)$.

\subsection{Quantum-mechanical model for false vacuum decay and analogy to Schwinger effect}\label{vac:QM}

\begin{figure}[h]
\centering
\includegraphics[scale=0.45]{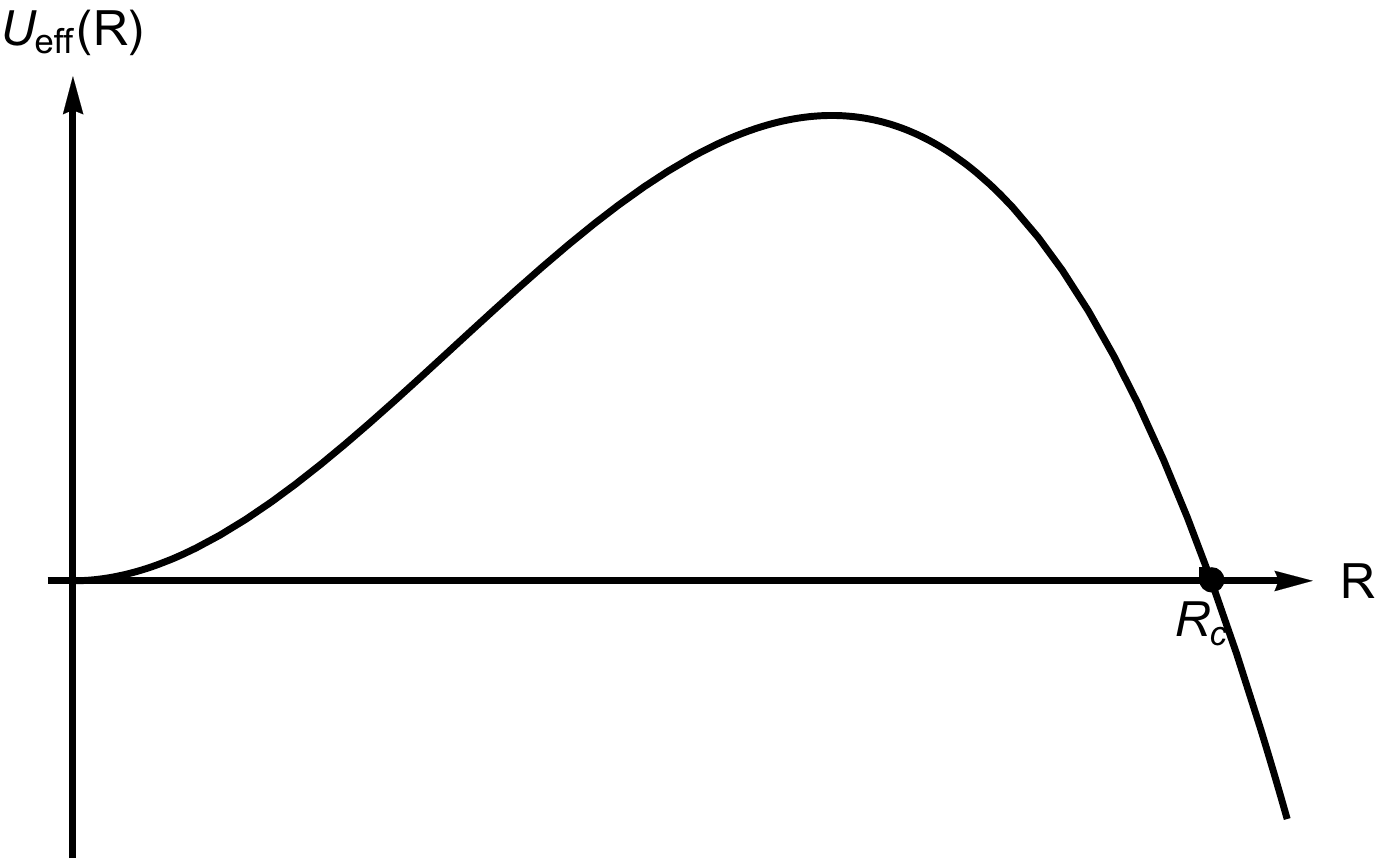}
\caption{The effective potential $U_{\rm eff}(R)$ for the bubble wall.}
\label{fig:bubblewallpotential}
\end{figure}

In the thin-wall regime, we can practically characterize the energy of the kink by the bubble radius $R$ and surface tension $\sigma$.
We approximate $\sigma$ as constant and use $R(t)$ as the dynamic variable. 
For a static thin-wall bubble, the energy is 
\begin{align}
\label{eq:utilde}
{U}_{\rm eff}(R)=4\pi R^2\sigma-\frac{4\pi}{3}R^3\epsilon,
\end{align}
where the first term comes from the surface tension and the second term comes from the negative energy density inside the bubble. One can view ${U}_{\rm eff}(R)$ as the effective potential for the bubble, shown in Fig.~\ref{fig:bubblewallpotential}. 
If we interpret $R$ as the position of a particle, the transition from the false to the true vacuum can be viewed as a quantum mechanical tunneling problem in analogy to Schwinger effect in the previous section.
Bubble nucleation occurs when the volume-dependent gain in energy due to $\epsilon$ exceeds the surface-dependent energy due to the tension $\sigma$, {\it i.e.}, for sufficiently large bubbles. 
The minimal radius $R_c$ for which this can happen is analogous to the critical distance $x_c$ for which pair creation becomes energetically favorable.

We model the bubble as a 
particle with effective mass $m(R)=4\pi R^2\sigma$ in the potential \eqref{eq:utilde} that initially stays at the origin $R=0$ with vanishing total energy. Classically, it is stable. However, quantum mechanically, it can tunnel to the exit point marked as $R_c$ in Fig.~\ref{fig:bubblewallpotential}, followed by a motion away from the critical radius (bubble expansion). $R_c$ is determined by ${U}_{\rm eff}(R)=0$, giving us $R_c=3\sigma/\epsilon$, in agreement with the result from the field theory.
For a general moving bubble wall, energy conservation implies
\begin{align}
\label{conservationlaw}
\mathcal{E}=\frac{m(R)}{\sqrt{1-\dot{R}^2}}-\frac{4\pi}{3}R^3\epsilon=0.
\end{align}
Note that when $R(t)\equiv 0$, this equation is trivially satisfied. When $R$ varies from 0 to a number smaller than $R_c$, the LHS of Eq.~\eqref{conservationlaw} is  always positive and hence the conservation law cannot be satisfied. Thus, classically, the only solution is $R(t)\equiv 0$.

For $R\neq 0$, Eq.~\eqref{conservationlaw} reduces to 
\begin{align}
\label{EoM-thinwall-QM}
\dot{R}^2-1+R_c^2/R^2=0,
\end{align}
taking the same form as Eq.~\eqref{eq:eom-schwinger}. As did for Schwinger effect, we move to the Euclidean time $\tau=\i t$ and have
\begin{align}
\label{Eucconservationlaw}
\left(\frac{\d R}{\d \tau}\right)^2=-1+\frac{R_c^2}{R^2}
\end{align}
with the initial conditions $R(\tau=0)=R_c$ and $\dot{R}(\tau=0)=0$.  We then obtain the same solution as \eqref{eq:motion} 
\begin{align}
\label{motion2}
R(\tau)&=\sqrt{R_c^2-\tau^2}\ \ \ {\rm for}\ -R_c\leq\tau\leq R_c\;, \notag\\ 
R(\tau)&=0\ \ \ {\rm for\ others},
\end{align}
which is in agreement with the result derived from the field theory.

To obtain the decay rate, we note that Eq.~\eqref{eq:utilde} can be derived from the following Minkowskian action:
\begin{align}
\label{eq:thin-wallQM}
S=\int\d t\left[-m(R)\sqrt{1-\dot{R}^2}+\frac{4\pi}{3}R^3\epsilon\right].
\end{align}
Taking $t\rightarrow -\i\tau$ and $\i S\rightarrow -S_E$, we obtain the Euclidean action
\begin{align}
S_E=\int\d\tau\left[m(R)\sqrt{1+\left(\frac{\d R}{\d\tau}\right)^2}-\frac{4\pi}{3}R^3\epsilon\right].
\end{align}
Substituting the Euclidean motion~\eqref{motion2} into the above action, one gets $B=27\pi^2\sigma^4/(2\epsilon^3)$. One may also estimate the prefactor $A$ by including the contribution from the zero modes corresponding to spacetime-translation symmetries and dimensional analysis as we did for Schwinger effect, obtaining
\begin{align}
    A\approx \left(\frac{B}{2\pi}\right)^2\frac{1}{R_c^4}.
\end{align}

Extending the model~\eqref{eq:thin-wallQM} beyond the thin-wall approximation may be carried out by using the functional Schr\"{o}dinger  equation and reducing the infinite field degrees of freedom to one or multiple degrees of freedom in a proper way~\cite{Gervais:1977nv,Bitar:1978vx,Copeland:2007qf,Tye:2009rb,Darme:2019ubo,Michel:2019nwa}. 

\section{Conclusions and Discussions}
\label{sec:conclusions}

In this work, we established an intuitive picture in which both, pair nucleation through Schwinger effect and false vacuum decay,
can be mapped onto a quantum-mechanical tunneling problem for a {\it relativistic} particle in one dimension.  
This analogy is based on the well-known facts that the \emph{potential energy} in both problems is related to a single distance variable  (the separation $x$ between the members of the Schwinger pair or the radius $R$ of a nucleating bubble), 
and that there exists a critical distance $x_c, R_c$
beyond which the potential energy is lower than that of the zero distance configuration.
For Schwinger pair, $x_c$ corresponds to the critical distance where the potential energy of the particles in the external field exceeds their rest mass.
In the case of bubble nucleation, $R_c$ marks the critical radius for which the volume-dependent gain in energy exceeds the surface-dependent cost to make a bubble. 
The novelty of our approach lies in the observation  
that the transition rate in both cases can be described rather accurately by 
interpreting $x$ or $R$ as the position of a relativistic particle 
that tunnels from $x,R=0$ to $x,R = x_c,R_c$
if the \emph{kinetic energy} is described with a suitably defined effective mass. 
For the Schwinger pair, the effective mass is simply $2m$, i.e., the actual physical mass of the pair. 
For the vacuum decay, the mass of the effective particle is $4\pi R^2\sigma$ where 
$\sigma$ is the surface tension of the bubble wall. 
This simple picture makes the analogy between Schwinger effect and false vacuum decay in the thin wall regime very clear.
We expect that our approach can be generalized to more complicated situations, such as pair production in curved spacetime, or when considering non-Abelian fields.

\section*{ACKNOWLEDGMENTS}

We would like to thank Valerie Domcke and Bj\"orn Garbrecht for proofreading the paper and for their helpful comments. We also thank Oliver Gould for beneficial discussions.

\begin{appendix}
\renewcommand{\theequation}{\Alph{section}\arabic{equation}}
\setcounter{equation}{0}

\section{Callan-Coleman method on quantum tunneling}
\label{app:callan-coleman}

\begin{figure}[h]
  \centering
  \hspace{10pt}
  \includegraphics[scale=0.4]{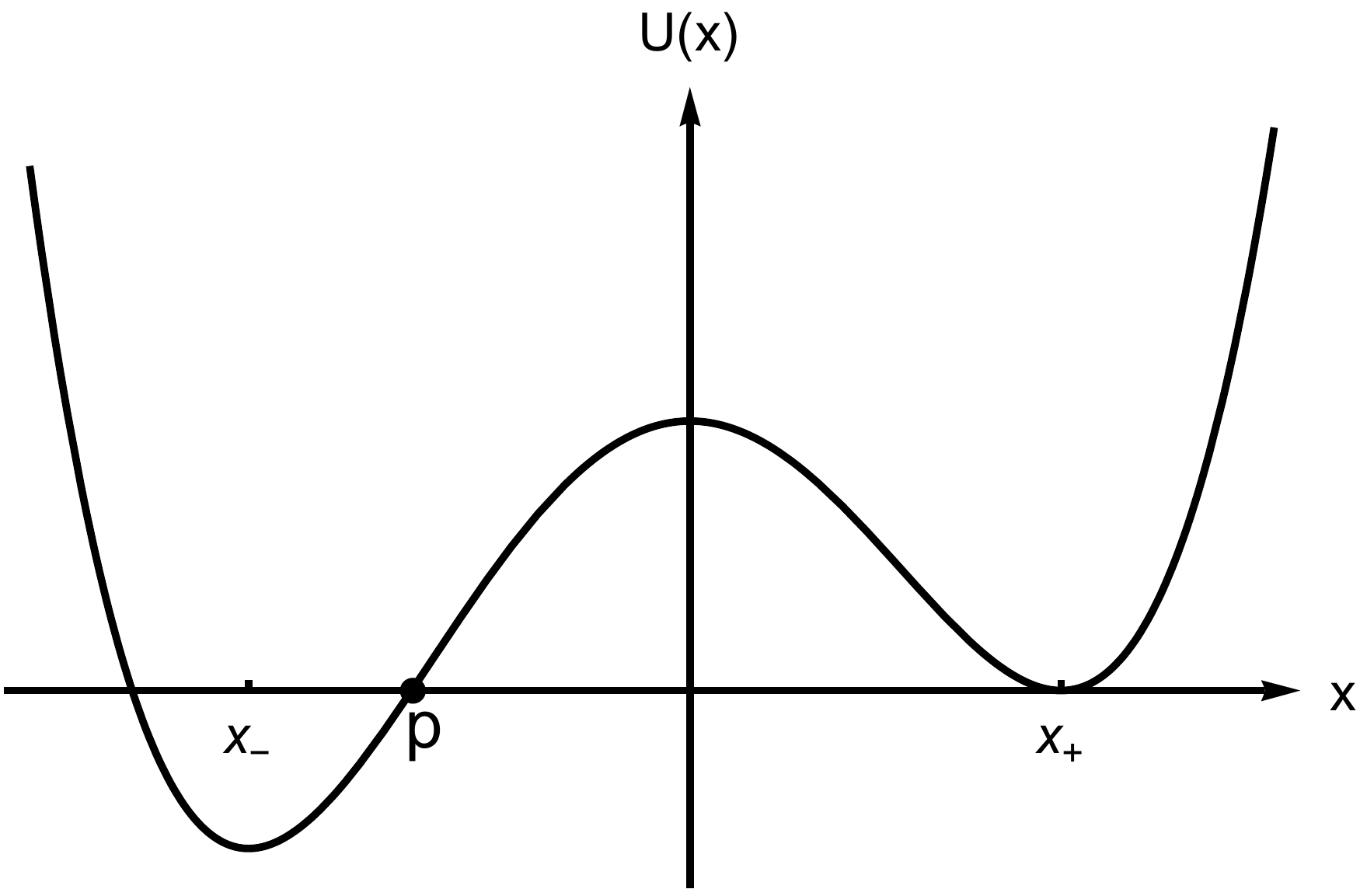}
  \caption{The classical potential $U(x)$ in a theory with a metastable minimum.
  \label{fig:potential2}}
\end{figure}

In this appendix, we briefly recall the Callan-Coleman formalism on quantum tunneling, closely following the original work~\cite{Callan:1977pt}. Suppose a particle initially occupies the ground state around the local minimum at $x_+$ shown in Fig.~\ref{fig:potential2}. While being stable classically, the particle can tunnel from local minimum $x_+$ through the barrier to the region around the global minimum $x_-$. 
In the figure, we also indicate the escape point $\mathbf{p}$ beyond which the motion of the particle can be described by a classically allowed trajectory. 

The tunneling rate 
can in principle be obtained by squaring the  transition amplitude
\begin{align}
\langle x_{\bf p}|\e^{-\i HT}|x_+\rangle=\int\D x\, \e^{\i S}.
\end{align}
Here $H$ is the Hamiltonian, $T$ is the amount of time during the transition  (typically taken to be infinity), and $S$ is the Minkowskian action. The path integral $\D x$ is performed over all trajectories that start at $x_+$ and end at $x_{\bf p}$.
A direct calculation of the tunneling transition amplitude is difficult because, for the boundary conditions of interest, there is no classical solution, i.e., no stationary point in the action that dominants the tunneling amplitude. So one cannot find a suitable way to perform perturbative expansion for the Minkowskian path integral.\footnote{It has been shown recently that one can still apply the method of steepest descent to the Minkowskian path integral for the quantum-tunneling problem by generalizing the contour integral in complex analysis to path integral~\cite{Ai:2019fri}. See Ref.~\cite{Fukushima:2019iiq} for a related discussion for Schwinger effect.} Fortunately, one can solve the problem in Euclidean space, using the Callan-Coleman method~\cite{Callan:1977pt}.

\begin{figure}[h]
  \centering
  \hspace{10pt}
  \includegraphics[scale=0.4]{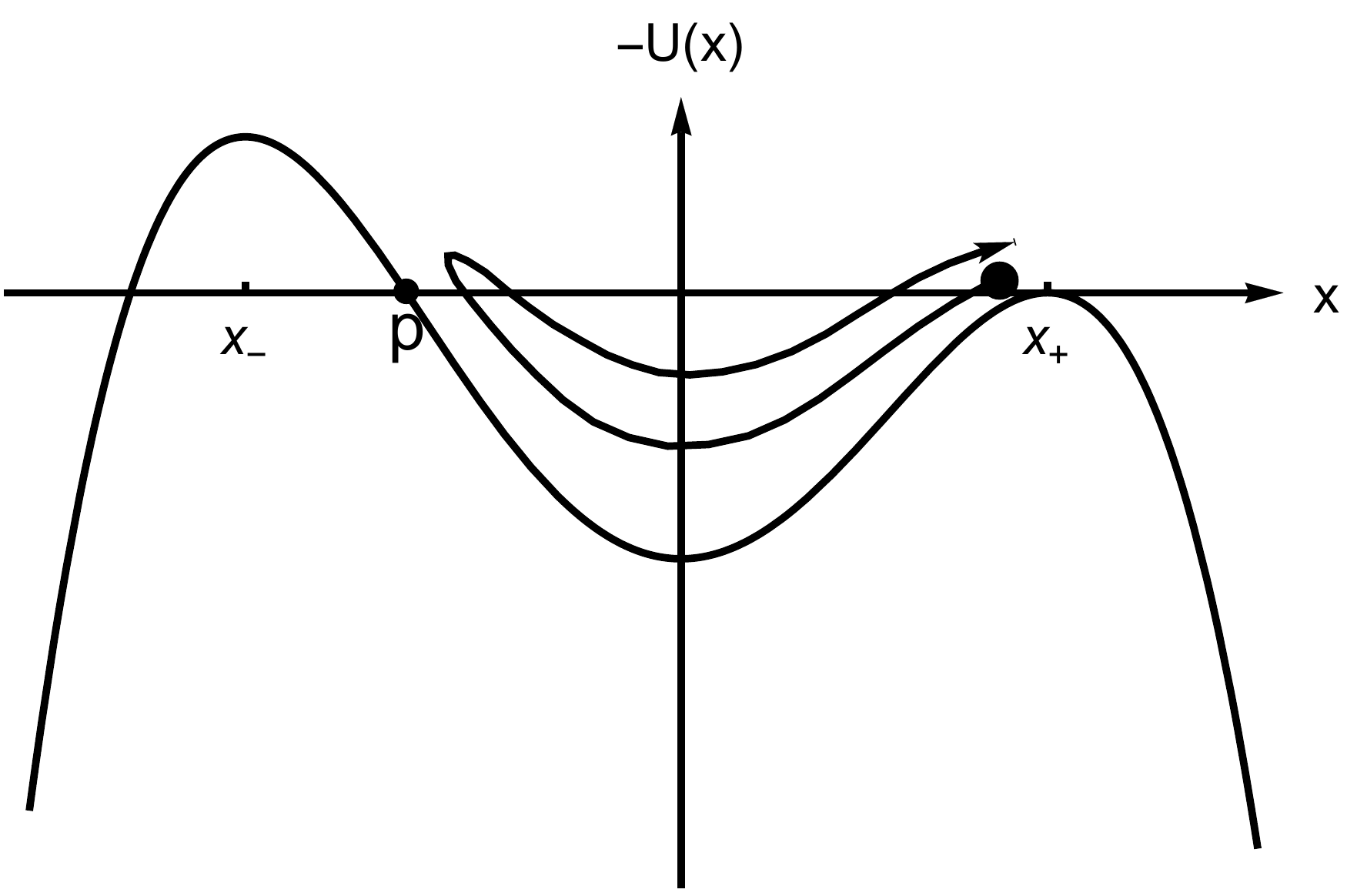}
  \caption{The potential is upside down in Euclidean space.
  \label{fig:bounce-ball}}
\end{figure}

Following Callan and Coleman, we instead consider the following Euclidean transition amplitude:
\begin{align}
\label{heatkernal}
\langle x_+|\e^{-H\mathcal{T}}|x_+\rangle=\int\mathcal{D}x\, \e^{-S_E[x]}\equiv Z^E[\mathcal{T}].
\end{align}
Inserting in the Euclidean transition amplitude a complete set of energy eigenstates, i.e.,
\begin{align}
\label{partition2}
\langle x_+|\e^{-H \mathcal{T}}|x_+\rangle = \sum\limits_n\,\e^{-E_n \mathcal{T}}\:\langle x_+|n\rangle\langle n| x_+\rangle,
\end{align}
and taking the large $\mathcal{T}$ limit, we thus obtain
\begin{align}\label{eq:E0}
E_0=-\lim_{\mathcal{T}\rightarrow\infty}\frac{1}{\mathcal{T}}\ln\left(\frac{Z^E[\mathcal{T}]}{|\langle x_+|0\rangle|^2}\right). 
\end{align}
Here $|0\rangle$ denotes the quantum mechanical ground state in the potential minimum around $x_+$, to be distinguished from the position eigenstate $|x_0\rangle$ at $x=0$ in Sec.~\ref{sec:schwinger}.\footnote{
A few comments on the intuitive prescription on extracting $\varGamma$ from the imaginary part of the ``energy" for the metastable state are in place. 
First, the false vacuum state $|0\rangle$ 
is not an eigenstate of the Hamiltonian and should not appear in the spectral resolution of the identity.
Second, the path integral in Eq.~\eqref{heatkernal} is apparently real and cannot give rise to a complex $E_0$ through Eq.~\eqref{eq:E0}. The reason why the Callan-Coleman method works out is subtle and has been carefully explained recently in Ref.~\citep{Ai:2019fri}.}
$E_0$ has an imaginary part 
which gives the decay rate as\footnote{As noted in Ref.~\cite{Ai:2019fri}, this formula describes the tunneling from the false ground state around $x_+$ to all possible final states so that the exit point of the tunneling is not necessarily $x_{\bf p}$. However, the tunneling from $x_+$ to $x_{\bf p}$ is the dominated process and this subtle difference is usually neglected.}
\begin{align}
\label{A6}
\varGamma=-2\;{\rm Im} E_0=\lim_{\mathcal{T}\rightarrow\infty}\frac{2}{\mathcal{T}}{\rm Im}(\ln Z^E[\mathcal{T}]).
\end{align}
Here we have used the fact that the amplitude squared does not contribute to the imaginary part.
One can evaluate Eq.~\eqref{heatkernal} through the method of steepest descent. We first need to identify all the stationary points in the path integral, i.e., the classical trajectories. In the Euclidean equation of motion, the potential is flipped upside down, cf.~Fig.~\ref{fig:bounce-ball}.
This allows for, besides the trivial solution $x_F(\tau)\equiv x_+ = \rm{const.}$ (which is called the false-vacuum solution in field theory), an instanton solution, named as bounce which starts at $x_+$ in the infinite past $\tau\to-\infty$, reaching the turning point $\mathbf p$ at a time $\tau=\tau_c$ known as collective coordinate of the bounce
and eventually bounces back to $x_+$ for $\tau\to\infty$, as shown in Fig.~\ref{fig:bounce-ball}. We denote the bounce solution as $x_B(\tau)$. In addition, there can be multiple  bounce solutions that also form stationary points.
In the so-called dilute-gas approximation, their impact on the path integral is approximated by a combination of $n$ subsequent bounces that are separated by time intervals much larger than the duration of each single bounce.

Expanding the Euclidean path integral around all the stationary points gives
\begin{align}
Z^E[\mathcal{T}]=Z^E_F[\mathcal{T}]+\sum_{n=1}^{\infty} Z_{B_n}^E[\mathcal{T}],
\end{align}
where the subscripts ``$F$'' and ``$B_n$'' indicate that the integral is evaluated by expanding $x(\tau)$ around the
``false-vacuum'' and $n$-bounce stationary points, i.e., $x(\tau)\equiv x_{F,B}(\tau)+\delta x(\tau)$.
In the dilute-gas approximation, the partition function $Z_{B_n}$ factorizes as
\begin{align}
    Z_{B_n}^E[\mathcal{T}]=Z_F^E[\mathcal{T}]\frac{1}{n!}\left(\frac{Z_B^E[\mathcal{T}]}{Z_F^E[\mathcal{T}]}\right)^n,
\end{align}
 where the appearance of $Z_F^E[\mathcal{T}]$ is due to the contribution from the trivial configurations between any two neighboring bounces. The factor $n!$ is due to the symmetry when exchanging the positions of the bounces in the $n$-bounce configuration. All the terms can be recollected into an exponential function, which eliminates the $\ln$ in \eqref{A6}.
One therefore finally finds the tunneling rate
\begin{align}
\label{eq:tunnelingrate}
\varGamma=\lim_{\mathcal{T}\rightarrow\infty}\frac{2}{\mathcal{T}}\left|{\rm Im}\, \left(\frac{Z^E_B[\mathcal{T}]}{Z^E_F[\mathcal{T}]}\right)\right|.
\end{align}
$Z_B^E$ is imaginary because the bounce is not a stable stationary point but a saddle point such that there is a negative mode in the fluctuations about the bounce. Taking the absolute value is due to a sign ambiguity when extracting the imaginary part. 
Expansion to second order in  $\delta x(\tau)$ gives the Gaussian approximation
\begin{align}
 \label{ZE_exp_formula}
    Z^E_{F,B}[\mathcal{T}]&\approx\int\D\delta x \, \e^{-S_E[x_{F,B}]-\frac{1}{2}\int\d\tau_1\d\tau_2\,\delta x(\tau_1)
    \G[x_{F,B}]
    \delta x(\tau_2)}\notag\\
    &\equiv A_{F,B}\,\e^{-S_E[x_{F,B}]},
\end{align}
where the quadratic fluctuation operator \eqref{eq:fluct-operator}
has to be evaluated at the false vacuum and the bounce, $
\G[x_{F,B}]\equiv\G(\tau_1,\tau_2)|_{x_{F,B}}$.
From Eqs.~\eqref{eq:tunnelingrate} and \eqref{ZE_exp_formula}, one can read off the semiclassical suppression factor  $B=S_E[x_B]-S_E[x_F]$ in the rate \eqref{eq:tunnelingrate2}.

The prefactor $A$ in the decay rate $\varGamma$ is determined by $A_{F,B}$ which can be expressed as the functional determinants of the operator \eqref{eq:fluct-operator} evaluated for fluctuations the bounce and false vacuum, i.e., $\G[x_{F,B}]$.
For the ``false vacuum,'' the valuation is straightforward and yields 
\begin{align}
    A_F=\left(\det G[x_F]\right)^{-1/2}.
\end{align}
For the bounce, however, there are several subtle points. First, the quadratic fluctuation operator evaluated at the bounce has a negative mode that originates from an unstable direction in \eqref{ZE_exp_formula}. The physical origin is the fact that the state $|0\rangle$ in Eq.~\eqref{partition2} is metastable, and this negative eigenvalue is the very reason why the imaginary part of $E_0$ in \eqref{eq:tunnelingrate} is nonzero.
Practically, this implies that the integral in \eqref{ZE_exp_formula} has to be solved by analytic continuation and the method of steepest descent to obtain a finite result.

Second, except for the negative mode, the quadratic fluctuation operator evaluated at the bounce also has zero modes. 
These can be related to symmetries in the action that are spontaneously broken by the bounce solution.
For example, the theories under consideration here are time translation invariant. 
This invariance is broken by the bounce solution that occurs at a specific time $\tau_c$, which we took to be zero before.
An infinitesimal shift of $\tau_c$ in $x_B(\tau-\tau_c)$ gives a different solution $x_B(\tau-\tau_c-\delta\tau_c)$. On the other hand, $\delta x(\tau)\equiv x_B(\tau-\tau_c-\delta\tau_c)-x_B(\tau-\tau_c)$ can be viewed as an infinitesimal fluctuation about the particular bounce $x_B(\tau-\tau_c)$. Since the action has time-translation symmetry, both solutions give the same classical action; $\delta x(\tau)$ therefore generates a flat direction in the fluctuations about $x_B(\tau-\tau_c)$ and incurs a zero mode for the corresponding fluctuation operator.
The integral over the zero modes can be traded for that over the collective coordinates with a Jacobian factor $\sqrt{B/2\pi}$ for each zero mode. For instance, in the example at hand, the zero mode corresponding to the time-translation symmetry gives a contribution $\mathcal{T}\sqrt{B/2\pi}$.
Each spatial dimension yields a similar factor, where $\mathcal{T}$ is replaced by the extension of spatial dimension. In a $d+1$-dimensional spacetime this  yields the overall factor\footnote{
Note that in quantum field theory, 
the $\varGamma$ in Eqs.~\eqref{A6} and~\eqref{eq:tunnelingrate} should be replaced by $\varGamma/V$. The volume factor $V$ from the zero modes is then eliminated by the denominator in $\varGamma/V$ and one ends up with Eq.~\eqref{FunctionalDeterminants} with $\sqrt{B/2\pi}$ replaced by $(\sqrt{B/2\pi})^{d+1}$.
}
\begin{eqnarray}
 \label{Aestimate}
V\mathcal{T}(\sqrt{B/2\pi})^{d+1},
\end{eqnarray}
where $V$ represents the volume of $d$-dimensional space. 
$\varGamma$ is then given by
\begin{eqnarray}
\label{FunctionalDeterminants}
\varGamma = \sqrt{\frac{B}{2\pi}}\left|
\frac{{\rm det'}\,\G[x_B]}{{\rm det}\,\G[x_F]}
\right|^{-1/2}\e^{-B}, \\
\notag
\end{eqnarray}
where a prime on $\det$ indicates that the zero modes should be excluded when evaluating the functional determinant.

\end{appendix}

\end{document}